# Recurrence Quantification Analysis and Principal Components in the Detection of Short Complex Signals

Joseph P. Zbilut

*Department of Molecular Biophysics and Physiology, Rush University, 1653 W. Congress, Chicago, IL 60612 USA; email, jzbilut@rush.edu*

Alessandro Giuliani

*Ist. Superiore di Sanita, TCE, V.le Regina Elena 299 Rome, 00161 Italy*

Charles L. Webber, Jr.

*Department of Physiology, Loyola University Chicago, Stritch School of Medicine, 2160 S. First Ave., Maywood, IL 60153 USA*

**Abstract**

Recurrence plots were introduced to help aid the detection of signals in complicated data series. This effort was furthered by the quantification of recurrence plot elements. We now demonstrate the utility of combining recurrence quantification analysis with principal components analysis to allow for a probabilistic evaluation for the presence of deterministic signals in relatively short data lengths. PACS: 05.40 05.45, 07.05Rm, 07.05Kf

*Key words:* recurrence quantification analysis, recurrence plot, nonlinear dynamics, principal components, complexity, chaos

## 1 Introduction

The need to determine that a given data sequence is a signal versus noise has engendered numerous algorithms, especially in the context of nonlinear or chaotic signals where traditional linear methods such as the Fourier transform are lacking [1-4]. Establishing this possibility is useful to prevent undue computational expense, and perhaps error prior to the employment of signal processing techniques. A relatively simple and straightforward graphical method, suitable for both linear and nonlinear signals, was introduced by Eckmann et al. [5]: place a dot at each position in an embedded distance array (time versus lagged values), which is approximately recurrent (Fig. 1). An important feature of such a graph, termed a recurrence plot (RP), is that line segments parallel to the main diagonal are points close to each other successively forward in time, and theoretically would not occur in a random as opposed to a deterministic process. While this feature can be visually well appreciated, the inability to perform hypothesis testing can, however, lessen its usefulness. As a result, several variables have been suggested to quantify RP's and have found utility in a wide range of scientific explorations [6-16].

Specifically, the following have been defined: the percentage of points that are recurrent (%REC – a global measure of recurrence); the percent of recurrent points which compose line segments and are therefore deterministic (%DET); the Shannon entropy of the histogram of varying line segment lengths as a rough measure of the information content of the trajectories (ENT); a measure of trajectory divergence derived from the length of the line segments which were claimed to be proportional to the inverse of the largest positive Liapunov exponent by Eckmann, et al. [5] (DIV); a least squares regression from the diagonal to the plot's corner as a measure of stationarity insofar as a flat slope indicates strong stationarity, whereas large slopes indicate poor stationarity due to changing values from one portion of the plot to another, i.e., a paling of the graph as originally suggested by Eckmann, et al. [5] (TREND); and mean distance (DIS) of the embedded points. Each of these variables provide information about different aspects of the plot, and are intercorrelated. Whereas use of individual variables such as REC or DET has been shown to be useful for quantifying signals, these variables in themselves do not give information regarding signal qualities such as divergence or transiency. Even though the use of the variable DET, can be justified for a preliminary inquiry as to the status of an unknown signal, random sequences can, in fact, also exhibit features of determinism simply by chance. Thus, it would be advantageous to determine which of the variables are most important for signal estimation. Since the variables are intercorrelated, their incorporation in a multiple regression would create an ill-conditioned model. A method which overcomes this problem is principal components analysis (PCA).

PCA is a well know data reduction technique based on singular value decomposition, and originally suggested by Pearson [17–18]. What distinguishes this method is that it does not require calculations in several embeddings or evaluation of fiducial points for trajectories, which may present ambiguous results – especially in cases where transients occur [1–4]. Furthermore, it allows for the "extraction" of significant features of the variables in a combination which allows for orthogonality (independence) of the components, while component loadings exhibit the relative weight of individual variables with respect to the given component [19–22]. The goal is to identify some component(s) based on a set of observed variables, which can discriminate between signal vs. noise.

## 2 Method and Example

Recurrence quantification analysis (RQA) was performed on a time series of typical signal processing and chaotic data; as well as randomly shuffled versions, and noise series to provide examples of nonsignals. All data were 1000 points long, with an embedding of 10, a delay of 32, a Euclidean norm for distance calculations, a neighborhood of 1 to define the recurrence, and line segments counted if composed of 2 or more points (Table 1)[1]. Signals were purposely set to be only 1000 points in order to mimic experimental reality, such as found under circumstances of nonstationarity especially in the biological sciences. Such an extreme restriction can make some signals, such as those derived from chaotic dynamics especially difficult to evaluate. The lack of adequate attractor sampling can create misinterpretations. For example, in the context of $D_2$ (correlation dimension) calculations, it has been estimated that the number of points necessary to calculate the dimension is dependent on the inequality $N \geq 10^{D_2/2}$, where $N$ is the data set length. But, for noisy or random data, higher embeddings may be necessary to overcome these effects, although, again because of limited experimental data length, this may be practically impossible if the delay method of reconstruction is used [23].

After submitting the data to RQA, the results were entered into PCA. Only principal component (PC) 1 (66.57 % variance explained) was found to be significant for a discriminant function (Table 2). Inspection of the loadings demonstrated large values for RQA variables DET, ENT, and DIV, clearly emphasizing deterministic (DET) as well as regularity features (ENT and DIV), although the remaining variables were also highly loaded, thus underscoring the importance of each variable for proper signal discrimination.

To test the utility of PC 1 for signal detection, it was used in a discriminant analysis (DA) for signal classification. This resulted in only 2 signals being misclassified (shuffled square wave, and the Hénon attractor) (Table 3). The misclassification of the shuffled square wave may be due to the relatively large DET value (32 %). This no doubt a result of shuffling the essentially long patches of constant data values which results in smaller patches, but nonetheless, are still present and contribute to DET (Fig. 2–3). In the case of the Hénon data, the calculated probability for inclusion as a nonsignal was marginal at 0.569; whereas the scrambled version was definitely excluded as a signal. Part of the problem is related to the DET value (9 %) for the scrambled version as opposed to the DET value of 8 % for the Hénon. Cearly, the other RQA variables were significant enough to exclude the scrambled version, but were marginal for the unscrambled version. In this respect, it should be noted that the discriminant function is based upon a simple binary decision; i.e., yes/no. Indeed, a third category of undecided could be defined based on probabilities in the range, e.g., from 0.4 to 0.7. Signals classified in this category would require further investigation.

---

[1] An embedding of 10 is routinely used to overcome some of the potential problems associated with data set length and noise; a delay of 32 was chosen based on the autocorrelation minimum for the sin signal, and continued for the others to maintain a constant set of data points. The literature suggests that a "correct" delay may be elusive. See ref. 23–24.

In order to determine if this PC extraction could correctly identify other complex signals, 1000 points of the logistic equation in the chaotic regime, the Lorenz attractor experiencing chaotic transients and crisis [25], and random numbers derived from beta radioactive decay [26] (in an effort to compensate from possible effects of pseudo-random number generators) were subjected to the DA as a test set (i.e., using the parameters established by the original training data set). By doing so, the results are subjected to less bias than in classifying the original set, since a classification function can produce optimistic results when it is used to classify the same cases that were used to compute it. The logistic equation was chosen since there has been some speculation that it may serve as a random number generator [27–28]. In all three cases, the data sets were correctly identified as signals (Table 4).

## 3 Conclusion

Use of RQA has demonstrated its utility in a wide variety scientific endeavors. The present results demonstrate its greater utility by combining the separate variables through PCA to provide a statistical estimation of signal probability. In combination with other more traditional forms of signal analysis such as FFT's, this approach may provide a useful adjunct. It is suggested that other forms of data compression such as neural nets may also be used in a similar fashion. Additionally, although not discussed here, the procedure may have relevance to evaluation of "complexity," since, as has been frequently mentioned, no one measure of complexity may be adequate [29].

# Software

Software developed by CLW and JPZ used in the analysis is available at http://homepages.luc.edu/ ˜cwebber in zipped format for DOS machines. It includes sample data files and detailed instructions for use.

# Acknowledgment

Grateful appreciation is expressed by JPZ to A. Giuliani and A. Colosimo of the University of Rome, "La Sapienza" Biochemistry Dept. for their hospitality during a Visting Professorship, at which time a portion of this work was completed.

# Figure Legends

Fig. 1. Recurrence plot (embedding = 3; delay = 1) for the logistic map ($\lambda = 4$). Note the short lines parallel to the main diagonal, which are inversely related to the largest positive Liapunov exponent.

Fig. 2. Square wave signal (top), and shuffled version (bottom). Although the shuffling destroys the regular phases of the square wave, it preserves features of determinism insofar as the essential binary oscillatory character of the signal is preserved by short, irregularly spaced patches. Thus, it may be argued that the misclassification is specious, since the shuffling was insufficient to destroy the signal's deterministic properties.

Fig. 3. Recurrence plot of shuffled square wave. Note the dense pattern of short parallel line segments.

# Tables

| Table 1. Signals Analyzed |
| --- |
| sin |
| shuffled sin |
| square |
| shuffled square |
| sawtooth |
| shuffled sawtooth |
| Hénon ($\alpha = 1.4$, $\beta = 0.3$) |
| shuffled Hénon |
| Lorenz (transient chaos and crisis; $\sigma = 10$, $\rho = 22.4$, $\beta = 8/3$) |
| Duffing ($\epsilon = 0.25$, $\gamma = 0.3$, $\omega = 1.0$) |
| shuffled Duffing |
| logistic ($\lambda = 4$) |
| chirp |
| Gaussian noise |
| uniform noise |
| $\beta$ decay |

| Table 2. Factor Loadings | |
| --- | --- |
| | Factor 1 (P = 0.002) |
| Dis | -0.747 |
| Rec | 0.644 |
| Det | 0.957 |
| Ent | 0.919 |
| Div | -0.929 |
| Trend | -0.649 |

| Table 3. Probabilities for Original Signal Set (Likelihood Ratio Chi-Square Probability for Classification = 0.003) | | |
|---|---|---|
| Signal | Probability for Signal | Probability for Nonsignal |
| sin | 0.989 | 0.011 |
| shuffled sin | 0.006 | 0.994 |
| square | 0.997 | 0.003 |
| shuffled square | 0.707 | 0.293 |
| sawtooth | 0.994 | 0.006 |
| shuffled sawtooth | 0.008 | 0.992 |
| Duffing | 0.996 | 0.004 |
| shuffled Duffing | 0.354 | 0.646 |
| Hénon | 0.431 | 0.569 |
| shuffled Hénon | 0.00 | 1.00 |
| chirp | 0.846 | 0.154 |
| Gaussian noise | 0.001 | 0.999 |
| uniform noise | 0.007 | 0.993 |

| Table 4. Probabilities for Test Set (Likelihood Ratio Chi-Square Probability for Classification = 0.051) | | |
|---|---|---|
| Signal | Probability for Signal | Probability for Nonsignal |
| Lorenz (transient chaos and crisis) | 0.999 | 0.001 |
| chaotic logistic | 0.772 | 0.228 |
| $\beta$ decay | 0.009 | 0.991 |

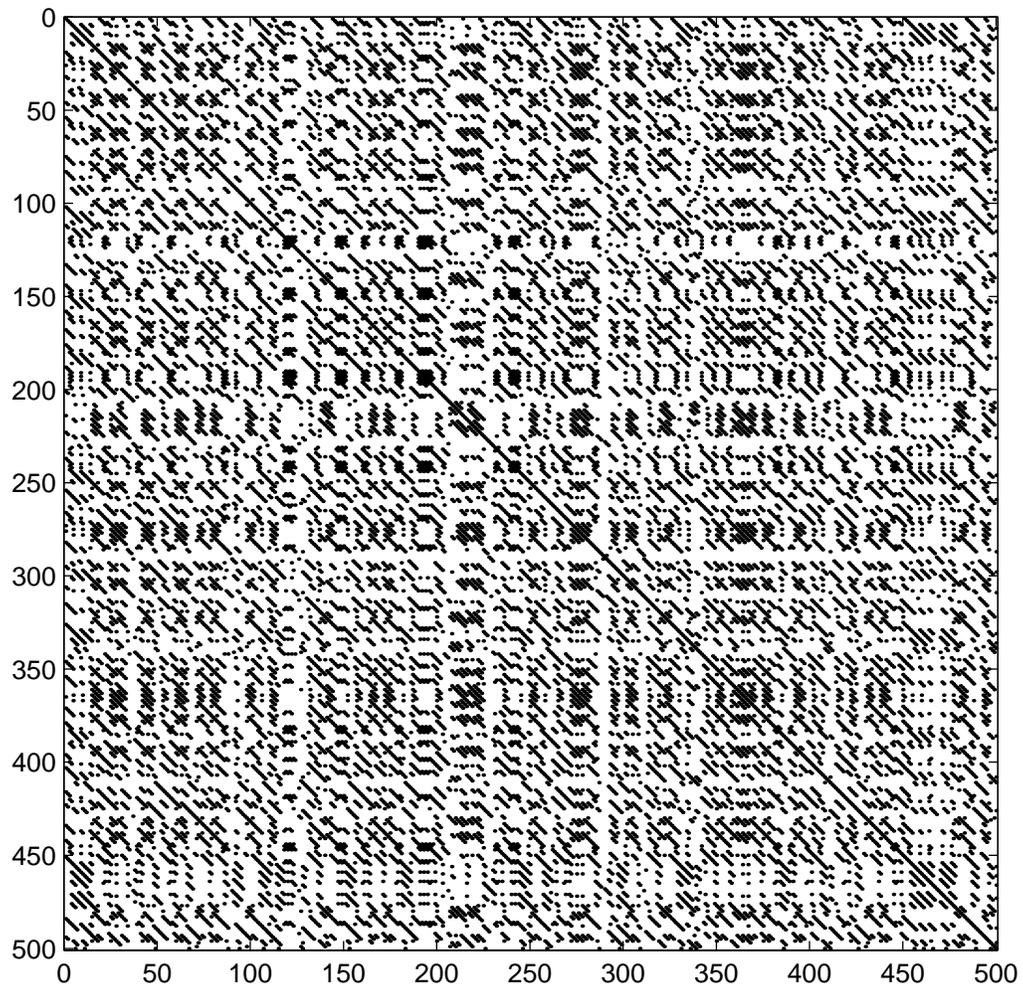

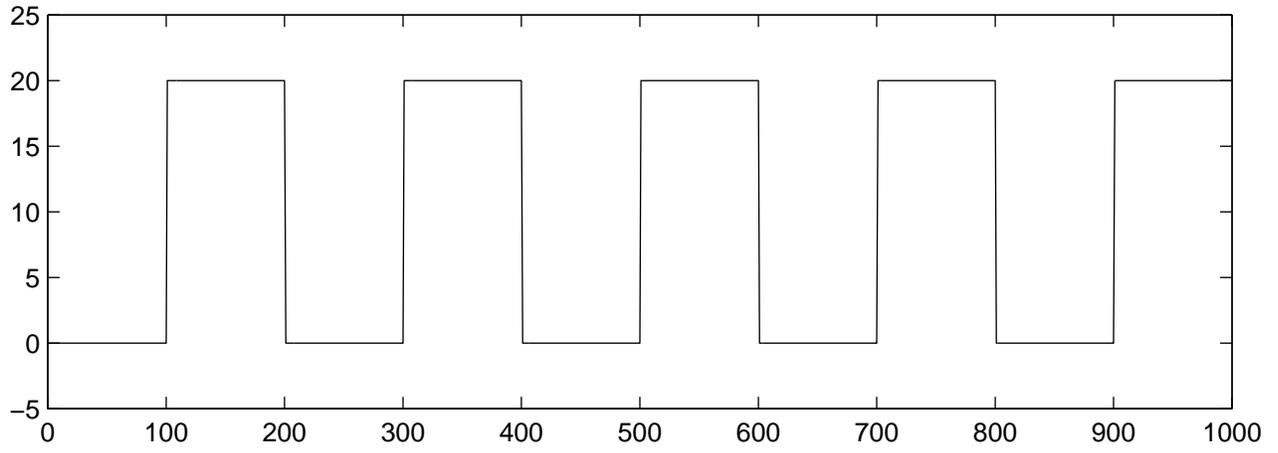
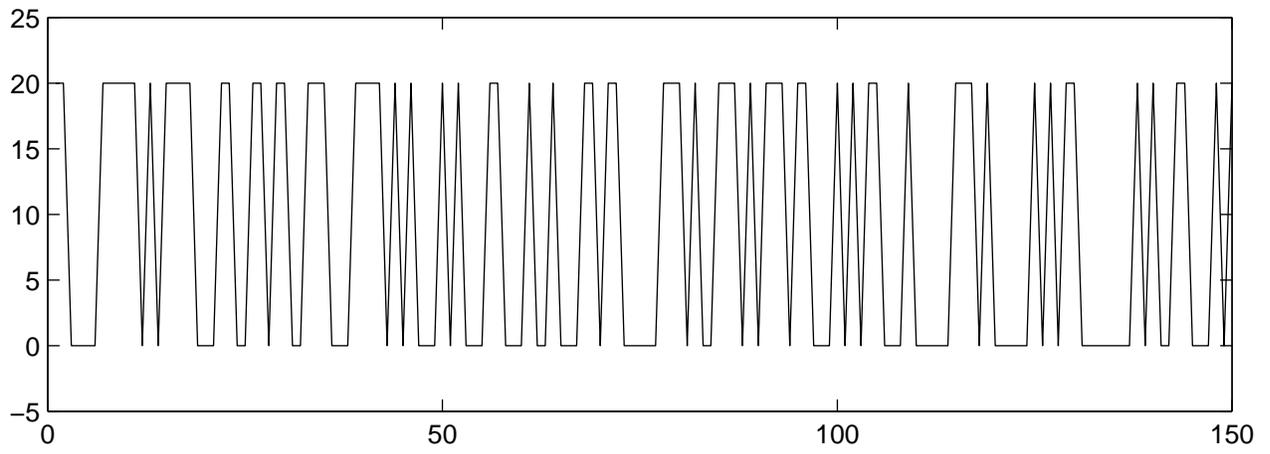

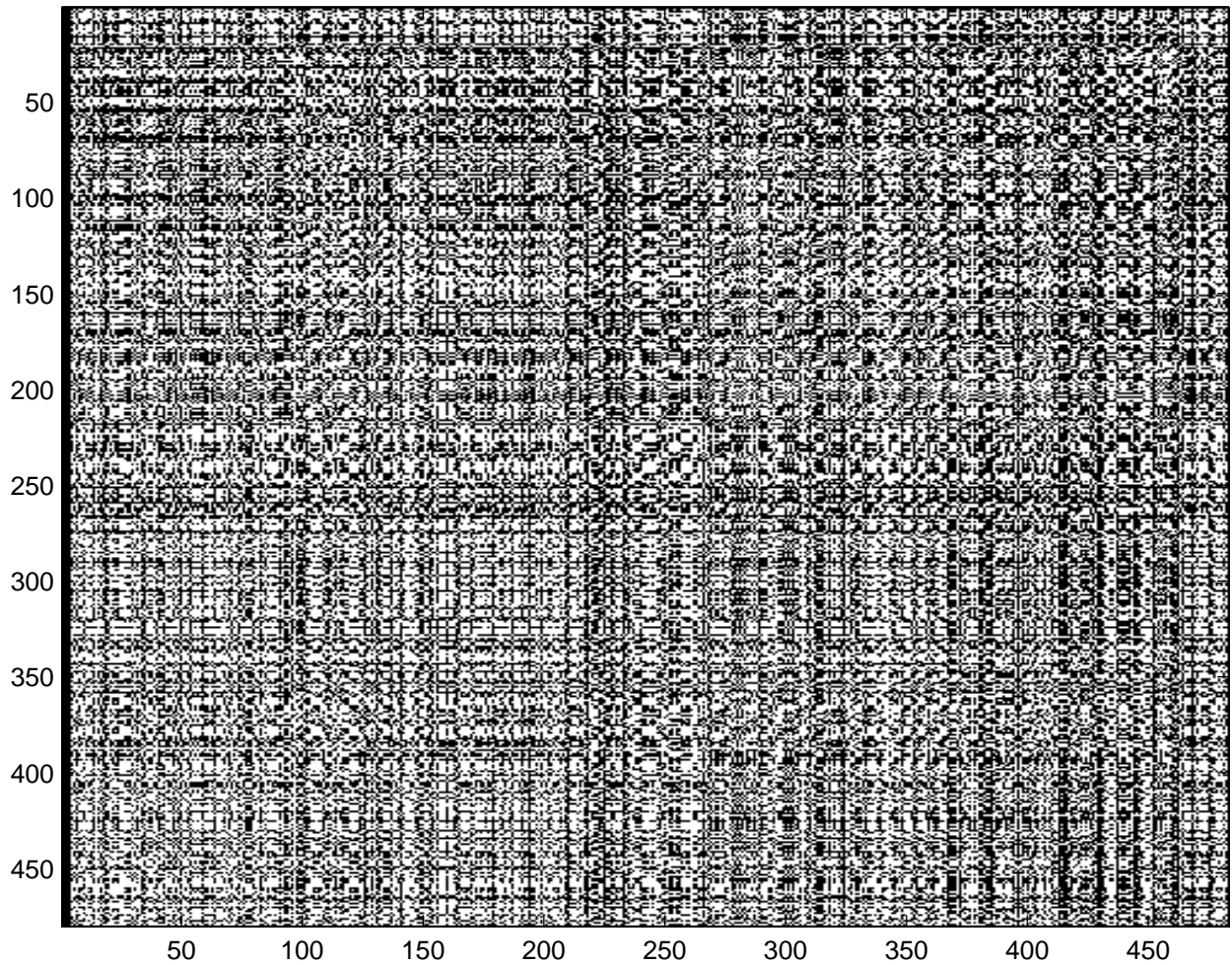